\documentclass[sigconf]{acmart}
\AtBeginDocument{%
  }

\copyrightyear{2026}
\acmYear{2026}
\setcopyright{cc}
\setcctype{by}
\acmConference[DAC '26]{63rd ACM/IEEE Design Automation Conference}{July 26--29, 2026}{Long Beach, CA, USA}
\acmBooktitle{63rd ACM/IEEE Design Automation Conference (DAC '26), July 26--29, 2026, Long Beach, CA, USA}
\acmDOI{10.1145/3770743.3804354}
\acmISBN{979-8-4007-2254-7/2026/07}

\usepackage{multirow}
\usepackage{subfigure}
\usepackage{hyperref}
\usepackage{acronym}
\usepackage{fancyhdr}

\fancypagestyle{FirstPage}{
\cfoot{
© Winklmann, Yu, Guo, Staudacher, Schulz {2026}. This is the author's version of the work. It is posted here for your personal use. Not for redistribution. The definitive Version of Record was published in {63rd ACM/IEEE Design Automation Conference (DAC '26)}, DOI 10.1145/3770743.3804354.}
}
\begin{document}

\acrodef{CPU}[CPU]{central processing unit}
\acrodef{FPGA}[FPGA]{field-programmable gate array}
\acrodef{GPU}[GPU]{graphics processing unit}
\acrodef{PSF}[PSF]{point-spread function}
\acrodef{NISQ}[NISQ]{noisy intermediate-scale quantum}
\acrodef{PL}[PL]{programmable logic}
\acrodef{PS}[PS]{processing system}
\acrodef{DRAM}[DRAM]{Dynamic Random Access Memory}
\acrodef{DDR}[DDR]{Double Data Rate}
\acrodef{IP}[IP]{Intellectual Property}
\acrodef{MMIO}[MMIO]{Memory-Mapped I/O}
\acrodef{HLS}[HLS]{High-level Synthesis}
\acrodef{std}[std]{standard deviation}
\acrodef{HPC}[HPC]{high performance computing}
\acrodef{API}[API]{application programming interface }
\acrodef{NAQC}[NAQC]{neutral atom quantum computer}
\acrodef{NA}[NA]{neutral atom}
\acrodef{BMFTR}[BMFTR]{Federal Ministry of Research, Technology and Space}
\acrodef{MQV}[MQV]{Munich Quantum Valley}
\title{Highly-Parallel Atom-Detection Accelerator for Tweezer-Based Neutral Atom Quantum Computers}

%
\author{Jonas Winklmann}
\orcid{0009-0009-4108-7732}
\affiliation{
  \institution{Technical University of Munich}
  \city{Garching}
  \country{Germany}}
\email{jonas.winklmann@tum.de}
\authornote{Both authors contributed equally to this research.}

\author{Yian Yu}
\authornotemark[1]
\affiliation{%
  \institution{Technical University of Munich}
  \city{Garching}
  \country{Germany}}
\email{yian.yu@campus.lmu.de}

\author{Xiaorang Guo}
\orcid{0000-0003-1697-817X}
\affiliation{%
  \institution{Technical University of Munich}
  \city{Garching}
  \country{Germany}}
\email{xiaorang.guo@tum.de}

\author{Korbinian Staudacher}
\affiliation{%
  \institution{Technical University of Munich}
  \city{Garching}
  \country{Germany}}
\email{staudacher@nm.ifi.lmu.de}

\author{Martin Schulz}
\orcid{0000-0001-9013-435X}
\affiliation{%
  \institution{Technical University of Munich}
  \city{Garching}
  \country{Germany}}
\email{martin.w.j.schulz@tum.de}


\begin{abstract}
Neutral atom quantum computers (NAQCs) are among the most promising computational platforms for quantum computing. Controlling and measuring individual atoms and their states, which often requires multiple imaging and image-analysis procedures, is typically the most time-consuming task during computation and contributes significantly to overall cycle times. 
To resolve this challenge, we propose a highly-parallel atom-detection accelerator for tweezer-based NAQCs. Our design builds on an existing state-reconstruction method and combines an algorithm-level optimization with a Field Programmable Gate Array (FPGA) implementation to maximize parallelism and reduce the run time of the image-analysis process. We identify and overcome several challenges for an FPGA implementation, such as introducing a prefetching mechanism to improve scalability and customizing bus transfers to support large bandwidths.

Tested on a Xilinx UltraScale+ FPGA, our design can analyze a 256$\times$256-pixel fluorescence image in just 115 $\mu s$, achieving 34.9$\times$ and 6.3$\times$ speedups over the original and optimized CPU baseline, respectively. Moreover, our accelerator can maintain consistent resource utilization across various atom array sizes, contributing to the ongoing efforts toward scalable and fully integrated FPGA-based control systems for NAQCs. 


\end{abstract}






\maketitle

\thispagestyle{FirstPage}
\section{Introduction}
\label{section:intro}
The capability to resolve single atoms through microscopic imaging has significantly advanced experiments in the fields of many-body physics and quantum simulation \cite{Schlosser:1, Bakr:1, Haller:1, Sherson:1, Morgado:1}. With the advent of quantum computing and the growing interest in \acp{NA} as a computational platform, the established \ac{NA} techniques are challenged by the need to produce a fast and integrated solution to compete with other modalities such as superconducting qubits. 

\begin{figure*}[htbp]
  \centering
  \includegraphics[trim=2cm 7.2cm 2cm 7cm, clip, width=0.8\textwidth]{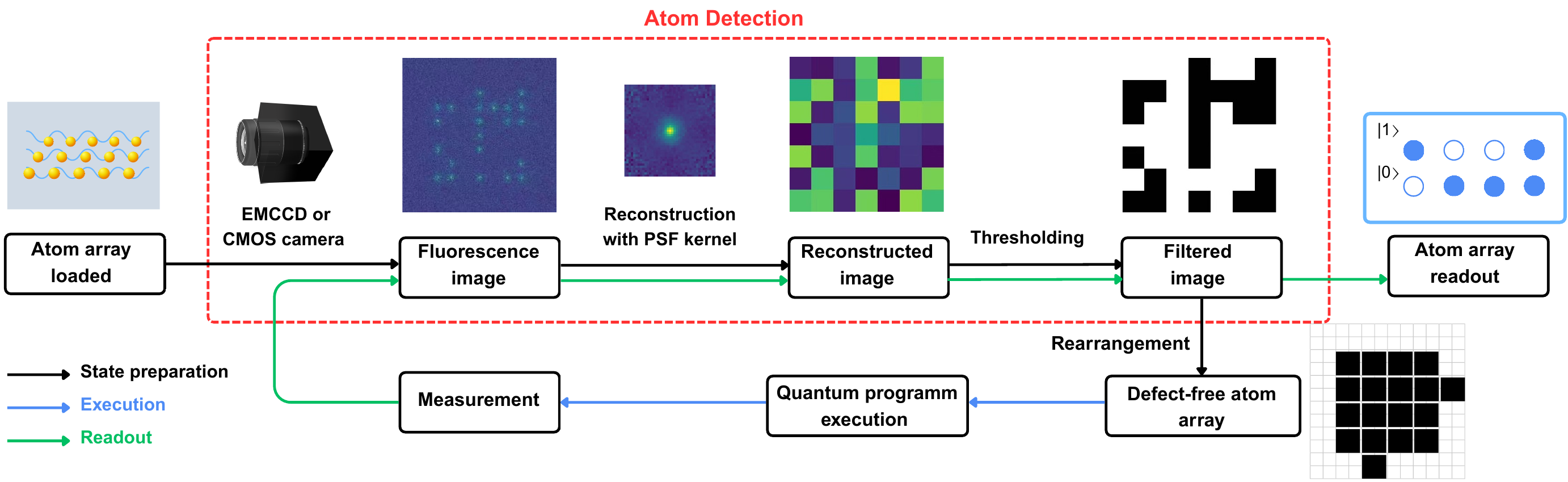}\\ 
  \caption{
  The overall workflow of a NISQ \ac{NAQC}, consisting of three main steps: 1) State preparation (black arrows), including atom detection and defect-free state preparation, 2) Quantum circuit execution (blue arrows), where quantum circuits are applied to the qubits followed by measurements, and 3) Final state readout (green arrows), involving atom detection process again to obtain the final qubit states. The atom detection, regarded as one of the most time-consuming operational steps, is illustrated by the dotted red line. 
  }
  \Description{Workflow of NAQC.}
  \label{fig:workflow}
\end{figure*}

Fig.~\ref{fig:workflow} shows the basic operation cycle of a \ac{NAQC} in the \ac{NISQ} era, where error-correction rounds are not considered~\cite{wintersperger2023neutral}. The cycle consists of four main stages. First, atoms are prepared and loaded into a two-dimensional optical tweezer array. Second, atom positions are detected through imaging, and the atoms are rearranged to form a defect-free atom array for computation. Third, quantum circuits are executed on the assembled array. Finally, the quantum state of each qubit is measured in the readout stage. As highlighted in Fig.~\ref{fig:workflow}, fluorescence imaging and state detection are already required twice in the \ac{NISQ}-level computation loop. This requirement becomes even more frequent with fault-tolerant quantum computing, where mid-circuit measurements must be performed repeatedly throughout deep quantum circuits~\cite{mid_FTQC}. Moreover, recent work indicates that mid-circuit readout should be pushed toward the "$\mu s$" scale to be capable of supporting multiple repeated rounds of measurement in practical quantum computers~\cite{bluvstein2024logical}. Therefore, \textbf{real-time atom detection is a key bottleneck in neutral-atom quantum computers.}




Moreover, between advancements in software efficiency and the development of parallel control hardware, there still exists a chasm. If one wants to progress \ac{NA} quantum computing to the point where traditional hardware developers would see it as a computer instead of a fast lab experiment, \textbf{an integrated control solution has to exist that is fast, robust, and independent enough not to rely on human intervention regularly}. In this case, \ac{FPGA}-based control architectures emerge as a promising solution, delivering a low-latency, programmable hardware required for real-time atom detection tasks.

Therefore, in this work, we propose \textit{a highly parallel atom-detection accelerator based on an optimized state-reconstruction approach}. Our solution utilizes the hardware-software co-design strategy first to optimize an existing atom detection method, namely the projection-based state-reconstruction algorithm~\cite{Wei:1}, to obtain a CPU-optimized version with high inherent parallelism and streamlined logic. Building on this, we then develop an \ac{FPGA}-based atom-detection accelerator that, once connected to a camera, can directly bridge the gap between image generation and \ac{FPGA}-based rearrangement \cite{Guo:1,wang2023accelerating} or readout procedures. Typically, the detection accelerator connects to the camera directly via a CoaXPress-capable FMC daughter card. However, since we focus on the algorithm and \ac{FPGA} designs in this work, we preload atom images into the \ac{DDR} memory as a replacement for live camera input. Moreover, with many microwave-pulse-generation platforms already being based on \ac{FPGA}~\cite{stefanazzi2022qick,liu2025risc,xu2023qubic}, the vision of a fully-integrated sorting and readout device is nearing completion.

We implement and evaluate the atom-detection accelerator on \ac{FPGA} with extensive experiments of different atom array sizes. As a result, our accelerator achieves up to 34.9$\times/$6.3$\times$ compared to the original implementation and our CPU-optimized implementation, respectively. Furthermore, the resource utilization remains stable for different atom array sizes due to our efficient prefetching mechanism. The results demonstrate a fast and scalable atom detection unit based on \ac{FPGA}, contributing to faster operation cycle of \acp{NAQC}. 

In summary, our main contributions include 
\begin{itemize}
    \item We optimize an existing state-reconstruction algorithm for atom detection, producing a CPU-optimized version tailored for parallel executions, which also enables a seamless extension to hardware accelerators.
    \item We develop an \ac{FPGA}-based accelerator for the optimized atom-detection algorithm, featuring a fully pipelined architecture with high parallelism to minimize the detection latency. 
    \item We implement the accelerator on a Xilinx UltraScale+ FPGA (ZCU216). Experimental results demonstrate that our accelerator efficiently handles various sizes of atom images, achieving a speedup of up to 34.9$\times$ and 6.3$\times$ over the original baseline and CPU-optimized version, respectively.
\end{itemize}

\section{Related Work}
\label{chapters:rlw}
\subsection{Atom Detection Algorithm}\label{chapters:atomDetectionAlgorithms}
Deconvolution algorithms have been employed extensively to solve atom detection tasks. 
\citeauthor{Rooij:1} compare several potential solutions, with Richardson-Lucy performing the best in terms of detection precision, closely followed by Wiener deconvolution \cite{Rooij:1}. Expanding on their algorithm repertoire, \citeauthor{Winklmann:1} introduce several more complicated fit-for-purpose algorithms like a global non-linear least-squares solver and \citeauthor{Wei:1}'s state-reconstruction library \cite{Wei:1}. They come to the conclusion that there is a clear tradeoff between precision and algorithm execution time \cite{Winklmann:1}. 

Out of the presented possibilities, we deem \citeauthor{Wei:1}'s projection-based state-reconstruction algorithm as the most suitable, as it performs very close to the overall most precise global solver while requiring a hugely decreased execution time. Originally aimed at state detection in lattices with much smaller spacings and drifting trap locations \cite{Wei:1}, some of its features, such as phase detection and individual projection kernels based on sub-pixel position, seem to be excessive for setups with well-separated atom locations. With the omission of these procedures, its perceived disadvantage of being conceptually comparatively complex dissolves, leaving behind only a number of element-wise matrix multiplications, as we will explain in Section~\ref{chapters:algorithm}. The fact that we deal with a straightforward calculation that is repeated once per atom site paves the way for a pipelined execution on the \ac{FPGA}.

\subsection{Deconvolution on FPGAs}
To the best of our knowledge, only a limited number of prior works have explored \ac{FPGA}-based deconvolution methods for reducing computational latency. Several Richardson–Lucy accelerators~\cite{Karine_RL_FPGA,RL_FPGA1,RL_FPGA2} have been proposed for motion and hyperspectral image reconstruction. However, due to the algorithm’s computational intensity, these approaches face scalability challenges on \acp{FPGA}, particularly with large kernel sizes or high-resolution input images. 

Notably, \citeauthor{bluvstein2024logical} use an FPGA-based qubit-decoding scheme that is developed with QuEra Computing. Their required decoding time is dwarfed by their exposure and image-readout time~\cite{bluvstein2024logical}. Unfortunately, they do not state the details of their analysis algorithm, and their solution is not public. Nevertheless, it is quite promising to see an integrated control solution that both industry and research entities are using.

As discussed in Section~\ref{chapters:atomDetectionAlgorithms}, there exists an algorithm with provably competitive precision that is well-suited to be adapted for usage on \acp{FPGA}. Therefore, in this paper, we build our detection accelerator based on this state-reconstruction algorithm, which primarily relies on convolution operations. In this case, we can benefit from resource multiplexing and parallel computation, achieving improvements in both latency and hardware efficiency. 
\section{Algorithm}\label{chapters:algorithm}
The spatial distribution of photons that are gathered by the camera can be represented as the convolution of the brightness at each atom location with the \ac{PSF}, which represents the response of the optical imaging system to a point-like light source or unit impulse~\cite{gualdron2024learning}. Since the resulting values are discretized into pixel values and as there are, assuming a negligible background illumination, only a certain number of point-like light sources, this convolution can be simplified as the sum of each atom site's \ac{PSF} multiplied by its brightness~\cite{Wei:1}.

\citeauthor{Winklmann:1} denote the expected number of photoelectrons registered at pixel $x$ as
$\lambda(x,\gamma) = b + \sum_{i=0}^nPSF_i(x) \cdot\gamma_i$ for a background illumination $b$ and $n$ atom locations, each of which being described by its \ac{PSF} $PSF_i(x)$ and brightness $\gamma_i$~\cite{Winklmann:1}.

The chosen state-reconstruction algorithm is based on the assumption that, since this equation is linear, we can find an inverse \ac{PSF} that will reconstruct the initial brightnesses from the pixel values of our fluorescence image. To do so, the Moore-Penrose inverse of the \ac{PSF} is calculated and, for each atom site, multiplied element-wise with the image details of equal size, centered around the location in question. The sum of this multiplication serves as the emission value, which, using a threshold, can be used to determine whether each site contains an atom or not~\cite{Wei:1}.
\subsection{Calibration}
As \citeauthor{Wei:1} states, the acquisition of the \ac{PSF}'s inverse is the most time-consuming aspect of the calculation~\cite{Wei:1}. However, the \ac{PSF} hardly changes unless the setup itself changes. As such, it does not need to be updated every time we run the calculation. Instead, we relocate it, along with other tasks that don't require execution during every cycle, to a preceding calibration stage where execution time is less critical.

Being fed an exemplary set of images, the calibration procedure automatically detects the position and angle of the atom grid, extracts the \ac{PSF} at each site, and calculates the inverse kernel and detection threshold. This calibration is performed offline and only infrequently, and it remains valid across many experimental shots (image frames).
\subsection{Runtime}
At runtime, we only execute those parts of the calculation that change non-negligibly for each compute cycle. Since the only variable that this applies to is the fluorescence image itself, the only calculation that requires execution at runtime is the element-wise multiplication with the projection kernel and subsequent summation and thresholding. 

\subsection{Optimization}
Since we want to produce a solution that is capable of being integrated into an \ac{FPGA}-based control system, it is not efficient to directly adopt the original algorithm~\cite{Wei:1}, which is inherently serial and algorithmically complicated. As such, we first follow the hardware–software co-design approach to reformulate and optimize the algorithm to make it execute efficiently on \acp{CPU} and, optionally, \acp{GPU}. Within our development, we observe that, since the sections of the program that are suited to run on \ac{GPU} are already comparatively fast and the transfer of data from \ac{CPU} to \ac{GPU} takes a non-negligible amount of time, an advantage of using this version could not consistently be established. As such, we will only use the \ac{CPU} version in the following chapters, referred to as the \ac{CPU}-baseline.

Due to the algorithm's focus on far smaller atom spacings, some aspects of it, such as the phase estimation, are superfluous or excessive for tweezer-based setups (the setup used in this work). Therefore, we refine it further to a tailor-fit solution for larger spacings running exclusively on \ac{CPU}. We also restructure the algorithm to expose inherent parallelism and simplify control flow. Instead of deploying parallel sections to a \ac{GPU}, we elect to employ OpenMP in order to utilize parallelization capabilities without suffering the drawback of transfer times between the two devices. We will refer to this version as the \ac{CPU}-optimized one, which is the foundation for the subsequent design of the \ac{FPGA} accelerator. It is to be noted that, while practically indistinguishable in their output for this use case, the \ac{CPU}-baseline and \ac{CPU}-optimized versions do not offer identical functionality, and any performance improvements do come at the cost of reduced adaptability to constraints such as smaller atom spacings and phase drift.
\section{FPGA implementation}
\label{chapters:impl}
In this section, we discuss the hardware design of our atom-detection accelerator, focusing on the customized reconstruction \ac{IP} as the core of this design. We first give an overview of the system architecture on the \ac{FPGA}, and then present the structure of the accelerator module in detail. 
\subsection{System Overview of Atom-Detection Accelerator}
\label{sec:sys_overview}
\begin{figure}[tb] 
  \centering
  \includegraphics[trim=17cm 7cm 16.8cm 7cm, clip, width=0.81\linewidth]{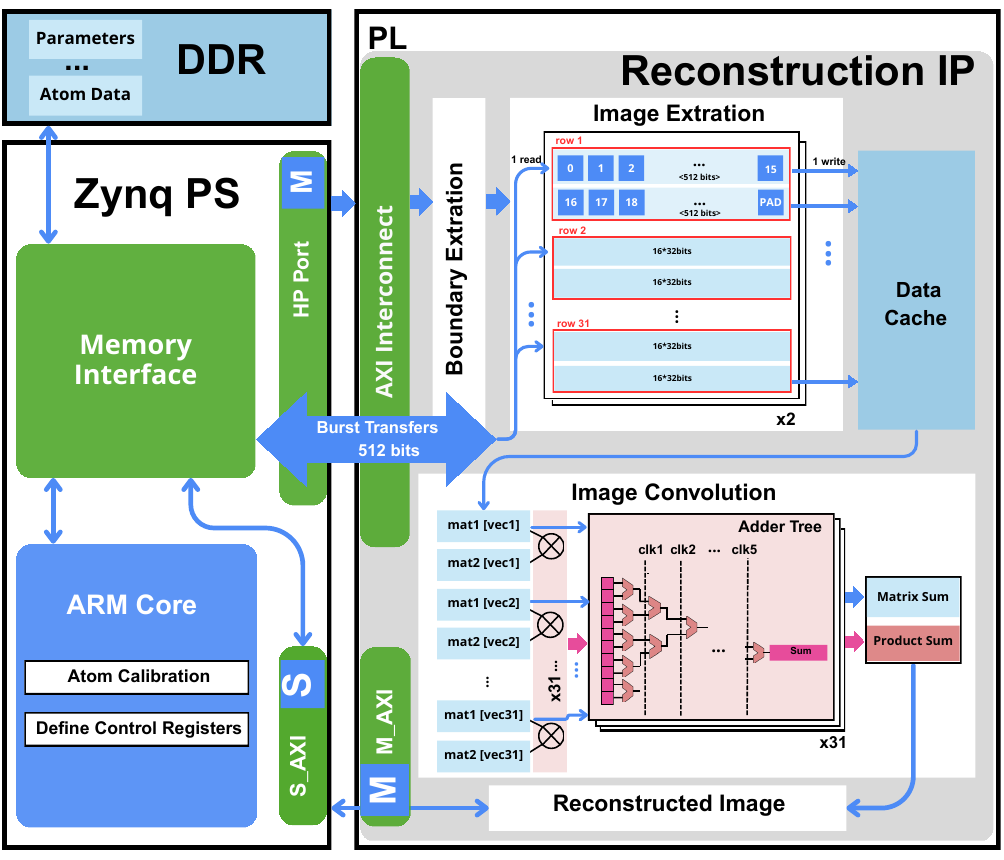}
  \caption{System architecture of the atom-detection accelerator, where the core is the reconstruction process to obtain the brightness of atoms. Here, \textit{mat1} corresponds to the projector derived from the PSF kernel, and \textit{mat2} corresponds to the atom's image.}
  \label{fig:design}
  \Description{System architecture.}
\end{figure}

Fig.~\ref{fig:design} illustrates the system overview of our FPGA-based accelerator, where the \ac{PL} and \ac{PS} work together to detect atom positions and apply the state-reconstruction algorithm. On the \ac{PS} side (ARM processor), primitive atom calibration is first implemented, which is mentioned as a mandatory step after we set up the system. Through this process, the approximate atom coordinates, fluorescence images, and convolution kernel information are stored in the \ac{DDR} memory. As noted before, storing the images in memory beforehand is only a temporary replacement for directly communicating with the camera via CoaXPress. All this data is then transmitted to the \ac{PL} side for logical processing via the AXI protocol, which features high-speed communication characteristics. Small data, like configuration files, is transmitted via the s\_axilite subordinate interface with low bandwidth, while reading or writing large amounts of data uses burst mode to enhance function throughput within a single request.

Although burst mode can intelligently aggregate memory accesses to the \ac{DDR} memory to maximize throughput~\cite{AMDhls}, transmitting only 32-bit data at one time can not utilize the full bandwidth of the AXI bus, which leads to longer data transmission time. Therefore, we first concatenate 16 instances of 32-bit data into a 512-bit vector, whose size is determined based on a trade-off between transmission latency and resource consumption in the following parallel processing part. After the data are transmitted to our customized logic on \ac{PL}, we split the 512-bit vector back to 32-bit pieces, and process them independently. 

We implement our accelerated reconstruction algorithm as a customized \ac{IP} core, which is controlled by the ARM processor through Python \ac{API}, where the \ac{PS} can access the control registers and address space of peripherals in \ac{PL} via \ac{MMIO}. Using this method, we can control the start and end of our accelerator. Finally, read the reconstructed image, i.e., obtain the emission values as output.


\subsection{Reconstruction IP}
Our dedicated hardware approach provides substantial benefit in the parallel execution of the state-reconstruction process. The entire detection accelerator adopts a dataflow design, as shown in the right part of Fig.~\ref{fig:design}, enabling task-level parallelism across four main modules: \textit{boundary extraction}, \textit{image extraction}, \textit{image convolution}, and \textit{output aggregation}.

Before the start of the accelerator, we already load the atom position grid, the \ac{PSF} kernel (we use 31$\times$31 in this work), along with the whole fluorescence image into the \ac{DDR} memory. For each atom in the atom array, the \textit{boundary extraction} module is responsible for getting the local image boundaries corresponding to its predefined coordinates. Based on the obtained indices, the \textit{image extraction} module fetches the corresponding image detail as well as the \ac{PSF} kernel from \ac{DDR} memory through the 512-bit AXI bus mentioned in Section~\ref{sec:sys_overview}. During this stage, the long 512-bit vector is decoded back to normal 32-bit data, and the \ac{PSF} kernel is processed into the so-called projector that serves as the input for the \textit{image convolution} module. 

To be noted, due to limited BRAM resources, storing all the data into BRAMs is infeasible, as the BRAM requirements will grow rapidly with the growth of the atom image size. To address this scalability problem, we employ a data cache in our accelerator, and we only need to load a single atom's pixel data and the projector at a time, thereby keeping the consumption of BRAM stable. Benefiting from the dataflow architecture, during the convolution time of the current atom pixels, the data for the next atom will be prefetched, and consequently, the scalability challenges can be eliminated.
Within the \textit{image convolution} module, two computations are performed in parallel: (1) we calculate the element-wise matrix multiplication between the atom's image detail and the projector kernel, obtaining the cumulative sum of this product, referred to as the \textit{product sum} (pink arrows in the \textit{image convolution} module in Fig.~\ref{fig:design}), and (2) we calculate the sum of all elements in the projector matrix, referred to as the \textit{matrix sum} (blue arrows in the \textit{image convolution} module in Fig.~\ref{fig:design}). During this computation, we employ a fully parallelized approach, where the matrix operations are decomposed into 31 concurrent vector processing units (defined by the size of \ac{PSF} kernel), each of which implements internal parallelization. Moreover, to reduce the latency of computing the sum, we deploy a logarithmic reduction algorithm to decompose the sum of a vector of 31 elements into four parallel stages, thereby building an adder tree structure. In this case, we can reduce the computational latency from $O(n^2)$ to $O(log (n))$, so we can finish the computation of adding 31 elements in five clock cycles. 

Furthermore, the original storage of the matrix data in BRAM restricts the parallelization factor due to the memory dependency problem, since we cannot read 31 values from the memory at one time. Therefore, we create a new vector space and fully partition the array to registers, thereby eliminating memory access conflicts and enabling simultaneous operations.

In the end, the \textit{output aggregation} module calculates a normalized value for each convolution result from the previous step based on equation~\ref{eq:output}, which provides the normalized brightness on each atom site. 
\begin{equation}
d_{\text{out}} = \sum_{i,j}(K[i,j] \cdot I[i,j]) \cdot \frac{\sum_{i,j}(K[i,j]\cdot u(i,j))}{\sum_{i,j} K[i,j]}
\label{eq:output}
\end{equation}
Where the sum of the element-wise multiplication of kernel $K$ with the local fluorescence image detail $I$ is multiplied with the fraction of the sum of used kernel values. $u(i,j)=1$ exactly if the kernel pixel at indices $i$ and $j$ fits into the image when the kernel is shifted to the investigated atom location, otherwise $u(i,j)=0$. This serves the purpose of normalizing the reconstructed emission for atom sites on the edges of the image where the full kernel does not fit. $\sum_{i,j}(K[i,j]\cdot u(i,j))$ is essentially the \textit{matrix sum} from before.
Afterwards, as described in Section~\ref{chapters:algorithm}, the normalized emission values are compared against a threshold, which determines the reconstructed state of each atom.

\section{Experiments and Evaluation}
\label{chapter:eva}
The evaluation of this work includes two aspects: specifically, 1) the result/output of the reconstruction algorithm, and 2) the run-time performance comparison across different platforms. Within the run-time comparison, we consider three test cases: a slightly optimized version of the original CPU-based implementation proposed in work~\cite{Wei:1} (CPU-baseline), our further optimized CPU version, slimmed down for tweezer-based setups (CPU-opt), and the FPGA-based accelerator. The CPU experiments are measured on an AMD EPYC 9374F 32-core processor, which is a model from the high-frequency line of AMD \acp{CPU} to ensure high sequential performance along with substantial parallelism. The \ac{FPGA} experiments are performed on a Xilinx UltraScale+ RFSoC ZCU216 board~\cite{AMDrfsoc}, which is programmed to operate at 100 MHz. We develop the accelerator in \ac{HLS}-compatible C++ and use Xilinx Vitis HLS 2024.2 to synthesize and package the customized IP core. The final system integration and implementation are carried out in Xilinx Vivado 2024.2. 
\begin{figure}[tb]
  \centering
  \subfigure[Raw image]{\includegraphics[width=0.3\linewidth]{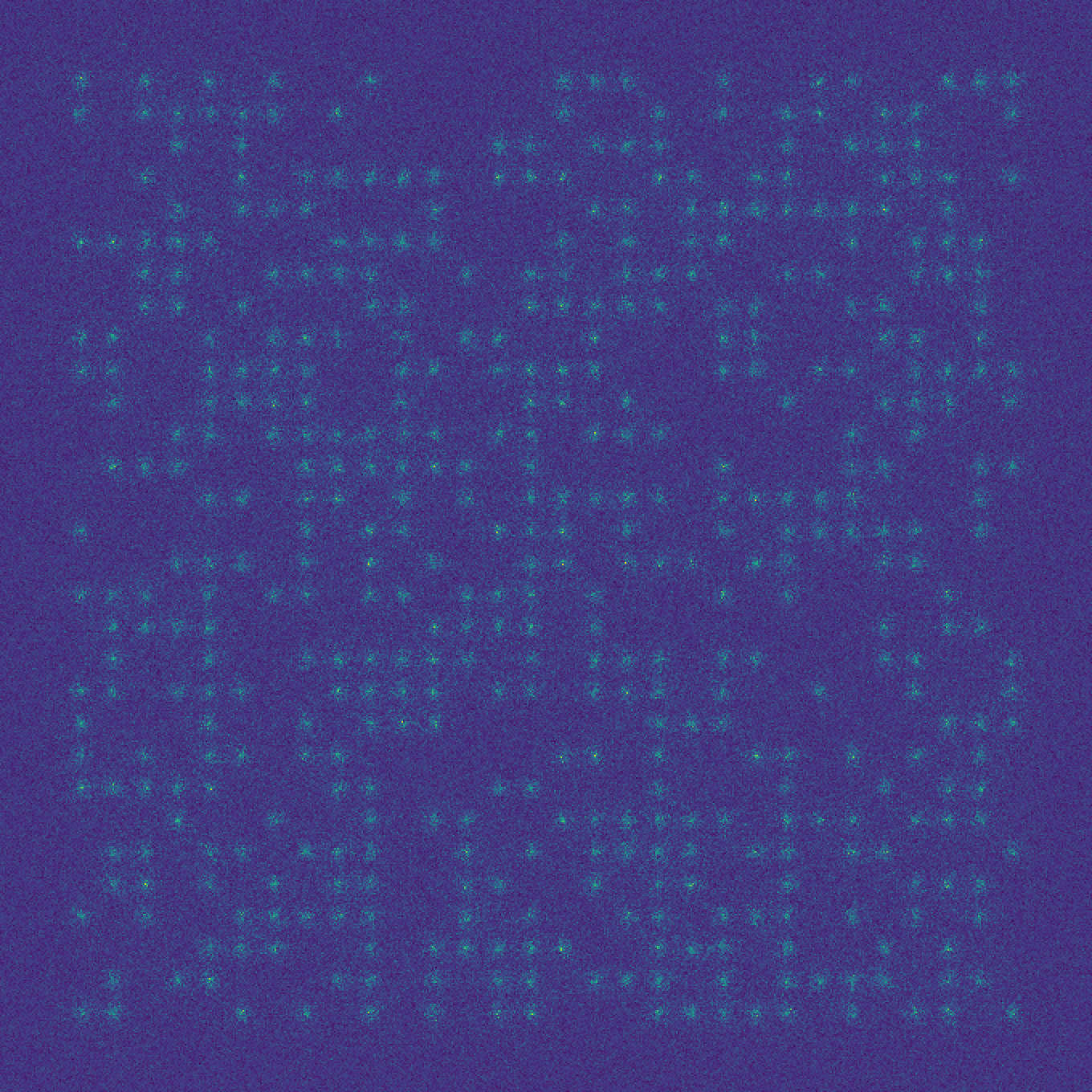}
  \label{fig:reconstruction_fig_camera}}
  \hfill
  \subfigure[Reconstructed image]{\includegraphics[width=0.3\linewidth,]{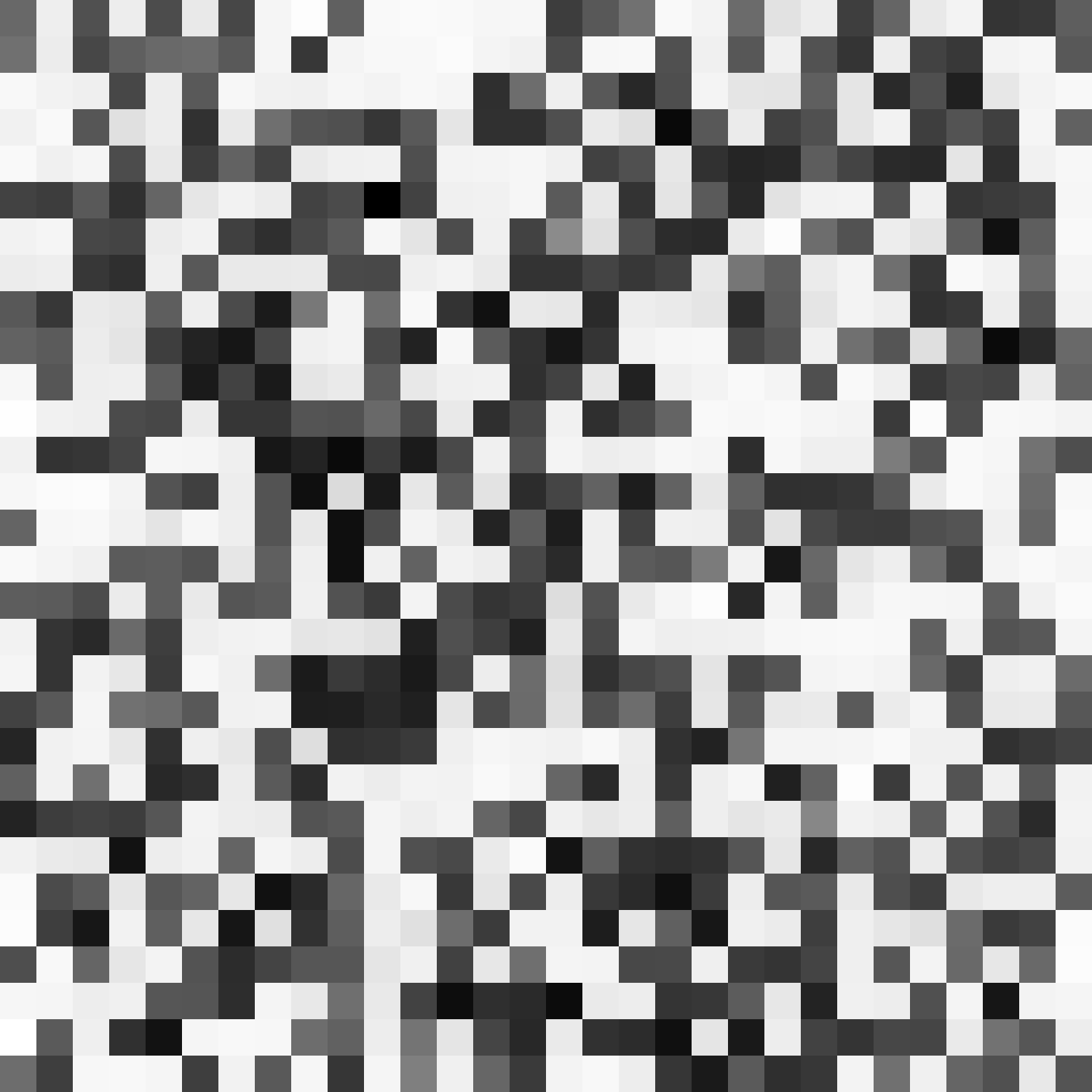}
  \label{fig:reconstruction_fig_recons}}
  \hfill
  \subfigure[Thresholded image]{\includegraphics[width=0.3\linewidth,]{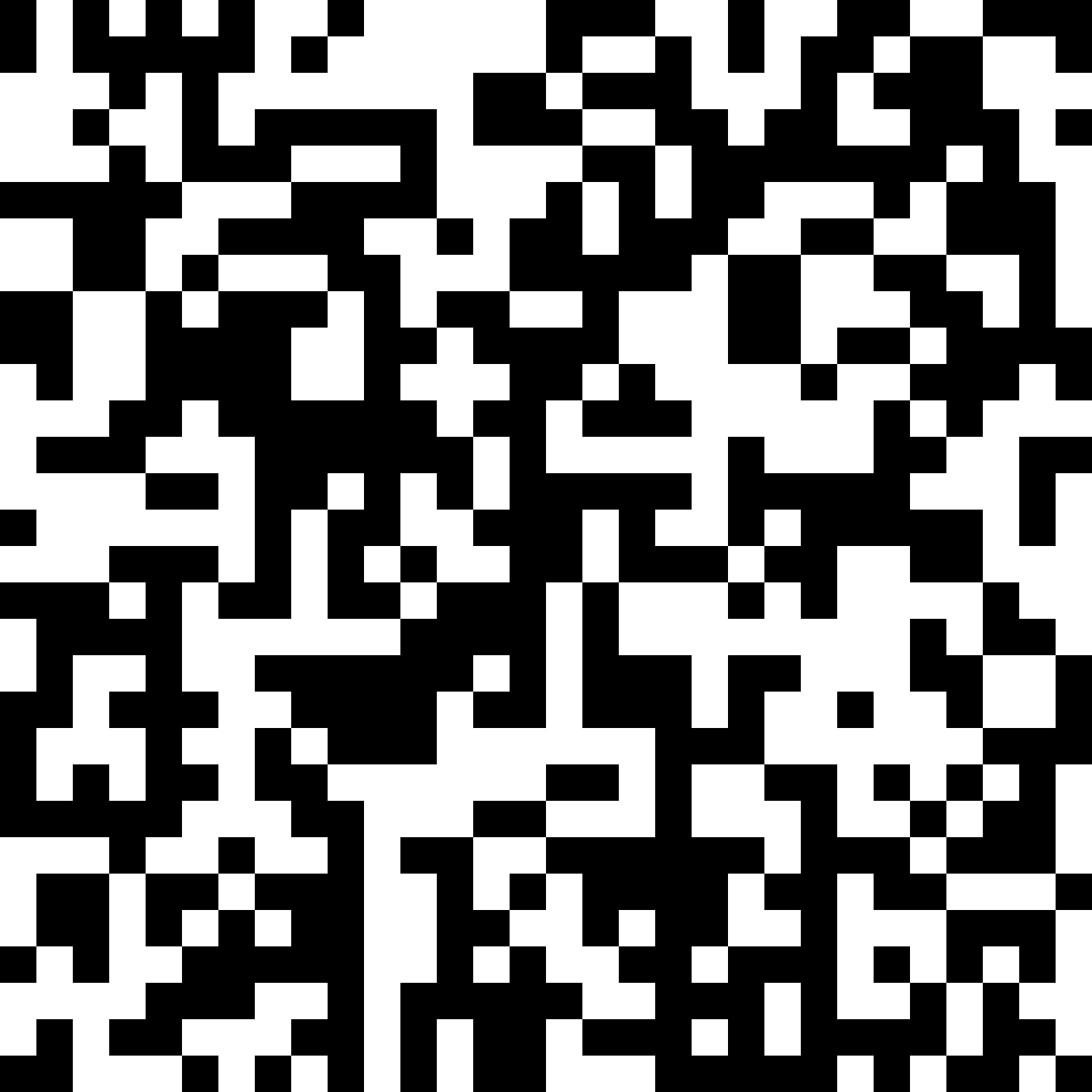}
  \label{fig:truth_fig_recons}}
  \caption{Result of the detection algorithm with a 30$\times$30 atom array as an example. (a) A raw image of the atom array captured by a camera (simulated). (b) The output of our reconstruction accelerator. Darker pixels denote higher reconstructed brightness. (c) Thresholded boolean result using a calibrated threshold. Black denotes a detected atom.}
  \label{fig:reconstruction_fig}
  \Description{Result of the detection algorithm}
\end{figure}

\subsection{Reconstructed Image}
Fig.~\ref{fig:reconstruction_fig} shows  the result of applying our atom detection algorithm to a 30$\times$30 atom array as an example. In this experiment, we firstly employ simulated images (see Fig.~\ref{fig:reconstruction_fig_camera} as one example) from work~\cite{simulator} in place of real camera images, as their quality is sufficiently similar for evaluation purposes. After the reconstruction procedure, we obtain a so-called \textit{emission matrix}, where its values of emissions indicate the brightness of atoms. Denoting higher emission values by darker pixels, we generate the reconstructed image of atoms shown in Fig.~\ref{fig:reconstruction_fig_recons}. After processing these \textit{emission values} with a predefined threshold, as described in Section~\ref{chapters:algorithm}, we will get the final '0' and '1' matrix as the final \textit{atom detection results}, which can be seen in Fig.~\ref{fig:truth_fig_recons}. We can use this boolean array in the subsequent atom rearrangement step.

Since the algorithm's precision has been shown previously \cite{Winklmann:1}, it is sufficient for us to ensure that our implementation's results do not deviate from the original ones. We have noted that minuscule differences occur that can be attributed to rounding errors.

\subsection{Run Time of Atom Detection}

\begin{figure}[tb]
  \centering
  \subfigure[Run Time]{\includegraphics[width=0.6\linewidth]{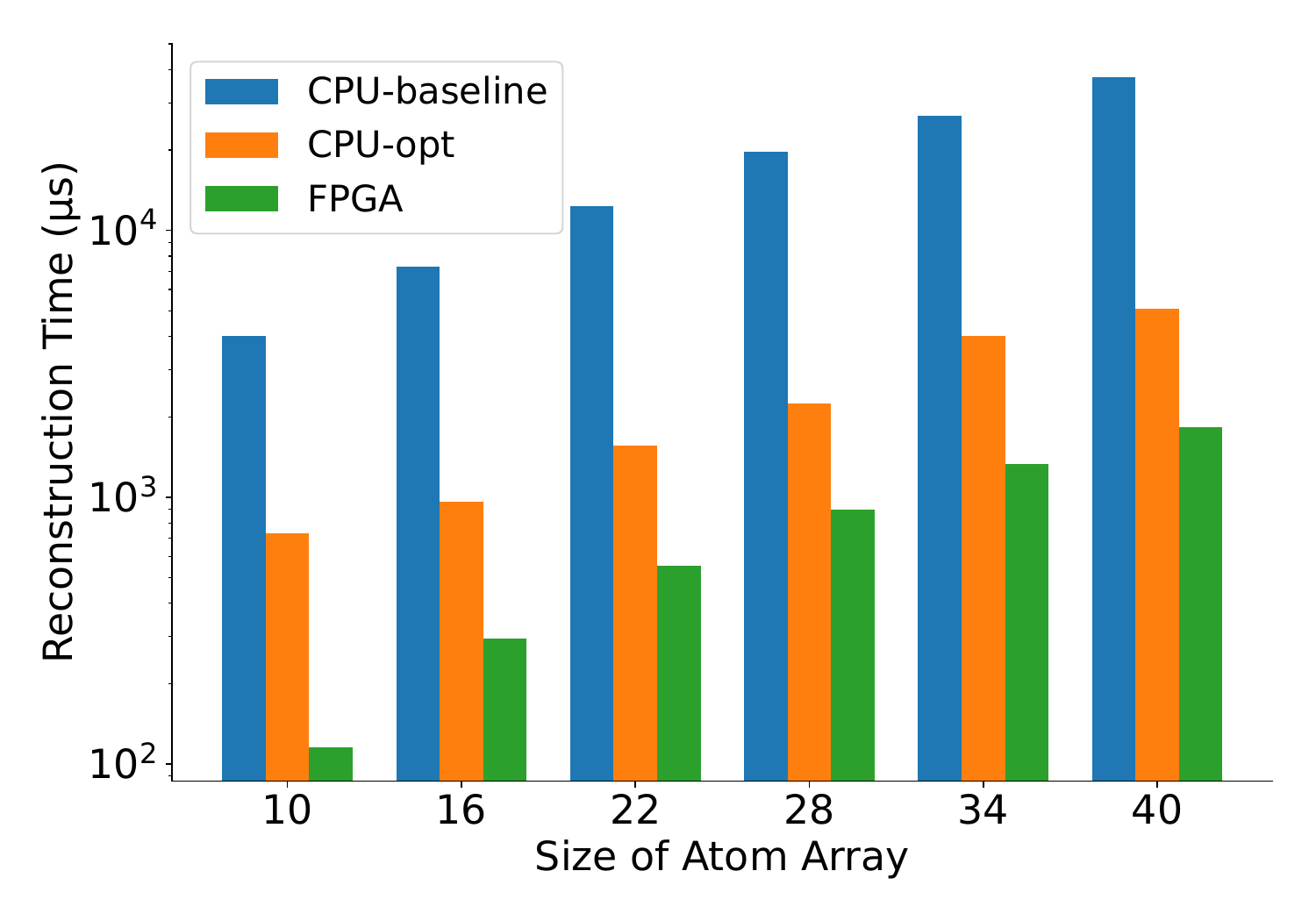}
  
  \label{fig:reconstruction_res:time}
  }
  \subfigure[Standard Deviation of Run Time]{\includegraphics[width=0.6\linewidth,]{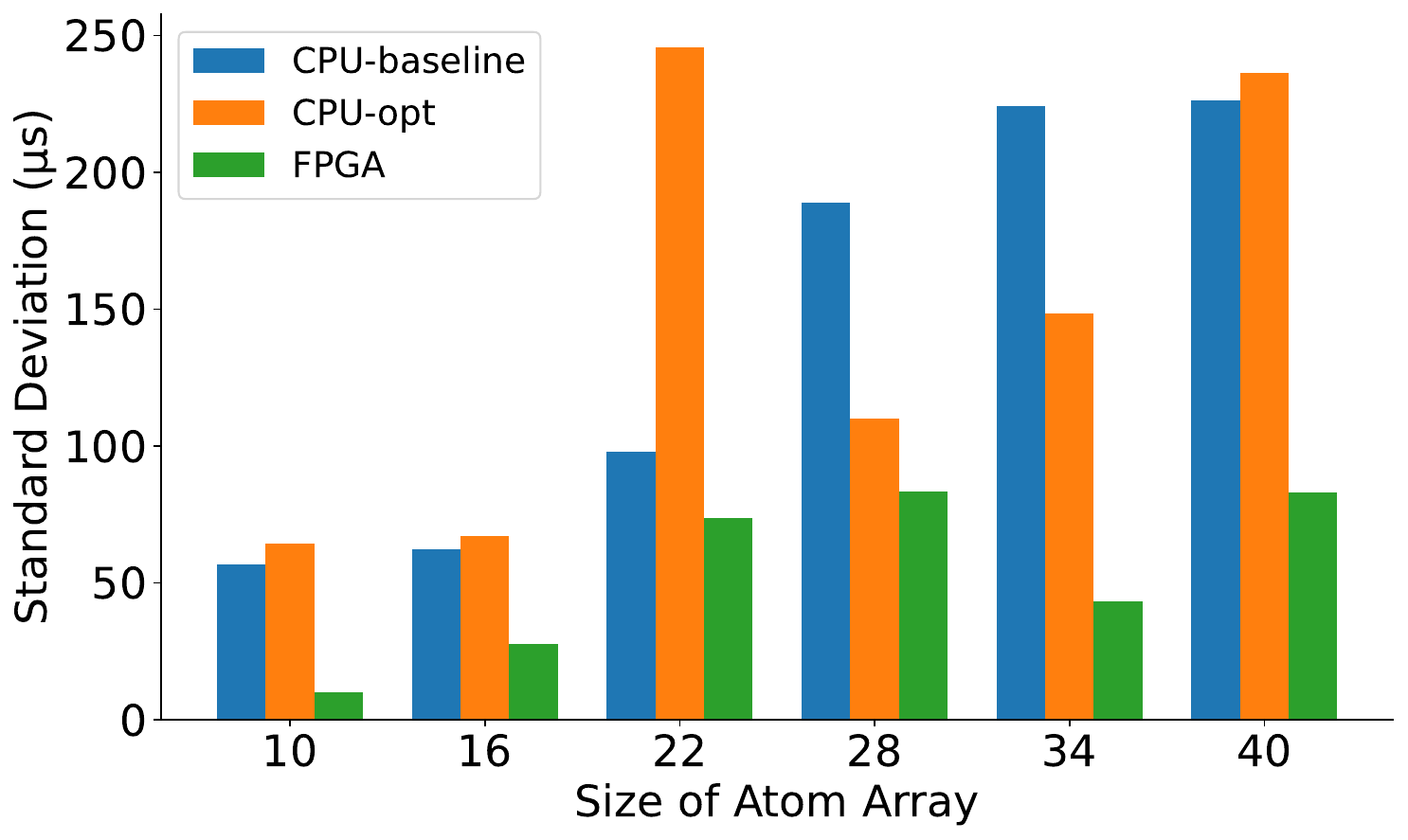}
  \label{fig:reconstruction_res:std}}
  \caption{Comparison of (a) run time of atom detection, (b) standard deviations of the run time for various atom array sizes (10$\times$10 to 40$\times$40) among CPU-baseline, CPU-opt, and FPGA. }
  \Description{Run time of atom detection}
\end{figure}

As shown in Fig.~\ref{fig:reconstruction_res:time}, in which the Y-axis is in logarithmic scale for better illustration, we compare the atom detection run time of the original algorithm, the CPU-optimized version, and the FPGA accelerator. The atom array size is varied from $10 \times 10$ to $40 \times 40$, corresponding to the image resolutions from $256 \times 256$ to $1024 \times 1024$ pixels. Within this experiment, we run each test 50 times, and report the average number as the final result. Across every atom array size, our FPGA accelerator consistently outperforms the other designs, but with acceleration gains decreasing as the array size increases. In more detail, for an array size of 10$\times$10, our FPGA-based solution achieves a speedup of 34.9$\times$ (115 $\mu$s vs. 4012 $\mu$s) and 6.3$\times$ (115 $\mu$s vs. 730 $\mu$s) compared to CPU-baseline and CPU-opt, respectively. For an array size of 40$\times$40, the speedup decreases to 20.6$\times$ (1.825 ms vs. 37.600 ms) and 2.8$\times$ (1.825 ms vs. 5.100 ms), yet the absolute reduction in run time is substantially larger. From Fig.~\ref{fig:reconstruction_res:time}, we also observe that this performance improvement comes from a two-step acceleration strategy following the hardware-software co-design strategy. First, CPU-opt already provides a partial acceleration over the baseline CPU implementation. Then, our FPGA-based accelerator delivers an additional speedup, resulting in a combined effect that maximizes overall performance.

Meanwhile, the FPGA accelerator also exhibits an advantage in run-time stability, as shown in Fig.~\ref{fig:reconstruction_res:std}. Although \acp{NAQC} benefit from their long coherence times, the deterministic execution time of each procedure can ensure that scheduled processes can be executed on time. To quantify this, we analyze the \ac{std} of the run time for CPU-baseline, CPU-opt, and FPGA versions across the 50 experimental runs. The results show that the FPGA accelerator can achieve a significantly more uniform run-time distribution compared to the two CPU versions, indicating a more stable performance. As both CPU-baseline and CPU-opt are executed on the same \ac{CPU}, their \ac{std} performance is generally comparable, yet notable variations also arise, which is consistent with the unpredictable behavior of the \ac{CPU}. In contrast, the accelerator is fully hardware-based, composed of gates and wires; therefore, the accelerator doesn't introduce intrinsic run-time variability. However, since the IP is controlled by the Python API on the ARM processor, issuing instructions and polling the control registers for time calculation both lead to the variability of the run time. Consequently, we can still observe a small \ac{std} in our experiments. But in the future, if the accelerator is integrated into the aforementioned fully \ac{FPGA}-based control system, this Python-related overhead and uncertainty could be eliminated eventually.


\subsection{Resource Utilization}

In addition to the run time of the detection algorithm, the resource utilization is another important metric to evaluate, especially in terms of scalability considerations. As introduced in Section~\ref{chapters:impl}, our architecture design employs a fixed parallelization parameter, including loop unroll factor and the parallel matrix process factor. Each calculation of a single atom and \ac{PSF} kernel will share the resources in a pipeline structure, featuring the time multiplexing property. Moreover, the size of the data cache also corresponds to the predefined \ac{PSF} kernel size. As a result, with the increase of the atom array size, the resource utilization remains the same, as shown in Table~\ref{tab:resource}, where the overall resource utilization is obtained from the Vivado implementation report. In this table, we also present the resource breakdown of each submodule. For all the resources except BRAM, the dominant contributor is the matrix convolution module, which is consistent with our design philosophy. This is expected, as the most computationally intensive operations are performed in this module, where we also apply the maximum parallelization factors. To be noted, as the Vivado synthesis/implementation report doesn't provide detailed utilization breakdowns for packed customized IP, these percentages shown in the breakdowns are estimated by the numbers in the Vitis synthesis report. 

{
\begin{table}[htbp]
\caption{FPGA Resource utilization and Breakdown across Modules}
\label{tab:resource}
\centering
\resizebox{\linewidth}{!}{%
\begin{tabular}{|c|c|c|c|c|}
\hline
\textbf{Module} & \textbf{LUT} & \textbf{FF} & \textbf{DSP} & \textbf{BRAM} \\
\hline
Total\footnotemark[1] & 109322 (25.71\%) & 131524 (15.46\%) & 447 (10.46\%)  & 67 (6.20\%)\\
\hline
\hline 
\multicolumn{5}{|c|}{\textbf{Breakdown by Submodules\footnotemark[2] ($\%$)}} \\
\hline
Boundary Extraction & 1.99\%  & 0.20\%    & 0.00\%      & 0.00\% \\
\hline
Image Extraction        & 0.59\%  & 0.51\%    & 0.00\%      & 0.00\%  \\
\hline
Image Convolution   & 22.03\%  & 13.62\%  & 10.15\%  & 0.00\% \\
\hline
Output Aggregation   & 0.10\%  & 0.06\%   & 0.31\%   & 0.00\% \\
\hline
Others                & 1\%     & 1.07\%   & 0.00\%      & 6.20\% \\
\hline
\end{tabular}}
{\raggedright\footnotesize\footnotemark[1]Obtained from Vivado implementation report.

\footnotemark[2]Estimated from Vitis synthesis report, since Vivado reports do not provide detailed utilization breakdowns for packed IPs.\par}
\end{table}
}

Overall, with the utilization of only one quarter of LUT, around 15\% of FF, and very low utilization of DSP and BRAM, this accelerator leaves enough space for other logic/components, making this design not only capable of being part of the fully integrated control system, but also part of the unified quantum control system, which is a promising solution to integrate quantum computing into \ac{HPC} system (HPCQC)~\cite{UQP,dobler2025survey,ramsauer2025towards}.

\section{Discussion}
\label{chapter:diss}
As demonstrated in this work, \ac{FPGA}-based atom detection accelerators represent a significant milestone in the state preparation and readout of \acp{NAQC}, providing opportunities for low-latency image analysis and mid-circuit measurement. Nevertheless, we do not claim that every lab and every experiment that images atoms requires an \ac{FPGA}-based solution. We see that many setups require flexibility and the possibility to intervene at any point. Especially experiments on high-density lattice-based atom arrays, three-dimensional structures, and other cutting-edge ideas, at this point, typically require the flexibility that only a software-only solution can provide. Instead, we envision our hardware solution as part of the control system for well-explored conditions where human intervention is obsolete. As \ac{NAQC} is right at the edge of being capable of offering value in production environments, we deem fully integrated control hardware as a necessity for the continued success of this modality.
\section{Conclusion}
\label{chapter:con}
This work presented the design and implementation of a highly-parallel state-reconstruction accelerator for the task of detecting atoms in \acp{NAQC}. Our design adopts a software-hardware co-design strategy, combining an algorithm-level optimization on the software side with an efficient FPGA implementation maximizing the computational parallelism on the hardware side. With this approach, our accelerator significantly reduces the computational run time of the reconstruction, contributing to a faster control pipeline. Experimental results demonstrate that our reconstruction accelerator achieves an ultra-low latency across all test image sizes, specifically 115 $\mu$s for a 10$\times$10 atom array and 1.825 ms for our largest testing image size of 40$\times$40. The result indicates up to 34.9$\times$ improvement compared to the CPU baseline, and 6.3$\times$ speedup compared to our optimized CPU design. Furthermore, the FPGA implementation maintains low hardware cost and exhibits excellent scalability. Overall, our work highlights the advancement of the FPGA-based image reconstruction accelerator and demonstrates its potential to be part of the FPGA-integrated control system for \acp{NAQC}. 
\begin{acks}
This work was funded by the German \ac{BMFTR} under the funding program \textit{Quantum Technologies - From Basic Research to Market} under contract numbers 13N16077 and 13N16087, as well as from the \ac{MQV}, which is supported by the Bavarian State Government with funds from the Hightech Agenda Bayern.
\end{acks}

\end{document}